# THE DEVELOPMENT OF A LAN FOR DVB-T TRANSMISSION AND DVB-S RECEPTION WITH DESIGNED QAM MODULATORS AND COFDM IN THE ISLAND OF MAURITIUS

Sheeba Armoogum[1], Vinaye Armoogum[2] and Jayprakash Gopaul[3]

[1]Department of Computer Science and Engineering, University of Mauritius, Reduit, Mauritius
`s.armoogum@uom.ac.mu`
[2]Department of Industrial Systems Engineering, University of Technology Mauritius, Port-Louis, Mauritius
`varmoogum@umail.utm.ac.mu`
[3]Mauritius Broadcasting Corporation, Reduit, Mauritius
`suraj.gopaul@yahoo.fr`

## ABSTRACT

*This paper is a thorough study of a digital television broadcasting system adapted to the small mountainous island of Mauritius. A digital TV LAN was designed with MPEG-2 signals. The compressed signals were transmitted using DVB-T and QAM modulators. QAM-16 and QAM-64 modulators were designed and tested with a simulator under critical conditions of AWGN and phase noises. Results obtained from simulation have shown that Digital video broadcast with a single frequency network (SFN) is possible in Mauritius with QAM-64 and QAM-16 modulators applying COFDM mode of transmission. However, this study has also shown that QAM-16 modulator had a better performance at low AWGN values (less than 12 dB) and can be adopted for Mauritius Island, provided that the number of transmitted channels is not high enough.*

## 1. INTRODUCTION

Today, the management of a broadcast station is impossible without the use of computer-based technologies. At the onset, these technologies merged the world of radio, television and telecommunication. Now, they are creating new values, new possibilities and new services. This means that these new tools are contributing to the growth of an 'information society', in the broadest sense of the term [1]. The real revolution is being instigated by the packetizing of digital programme signals and the distribution of these packets in the form of data streams over internal and external networks. This will have a tremendous impact on radio and television and will entirely change their future form. In the studio, new computer-based tools are allowing us to store our programme content on servers, and to access or distribute this content via IT-based networks [2]. All cumbersome hardware equipment is now being replaced by software, that is, by something that can easily be changed and updated over time. This aspect is extremely important. These mammoth changes are, of course, leading to completely new workflows, particularly in the broadcast environment [1].

In 1999, COFDM (Coded Orthogonal Frequency Division Multiplexing) technology was implemented, and it subsequently provided the first step in support of the migration to fully digital transmission system architecture. The implementation and modulator architecture facilitate the interface to most types of digital transport streams, e.g. asynchronous serial interface (ASI), which is used to carry a compressed digital video bit stream [3]. The advantages of COFDM technology are apparent in its ability to offer error free transmission under severe multi-path conditions and its ability to occupy less overall bandwidth than its previous analog FM counterpart [3,4,5]. A typical COFDM waveform occupies 8 MHz of bandwidth at its 1dB bandwidth points [3]. Its digital implementation is based on the ETSI standard EN300-744 for DVB framing, channel coding and modulation architecture [5]. The uniqueness of the standard allows the user to customize his data throughput as a function of three main parameters: forward error correction, guard interval, or delay spread and modulation type [6].

Following international trends in the way broadcasting technologies are evolving, the local broadcast station has no choice other than to walk on the paste of these changing technologies [7]. Maintenance of Beta Cam analogue equipment is becoming more and more costly as parts are becoming rare in the international market. However, swapping to new digital technologies is not an easy task. The implementation of new digital broadcasting techniques is being done phase by phase over some period of time. At the broadcasting station, new digital equipment has been installed in parallel to the old analogue equipment. Personnel are being trained to work on the new equipment. At this stage of digitalization, lots of problem is becoming apparent. There is no compatibility between the old format of transmission and the new one. Digitalization and networking in broadcasting house concerns four mains department (figure 1), namely the News production (editing), production department (editing), Archives and preview unit, Television transmission and satellite reception room.

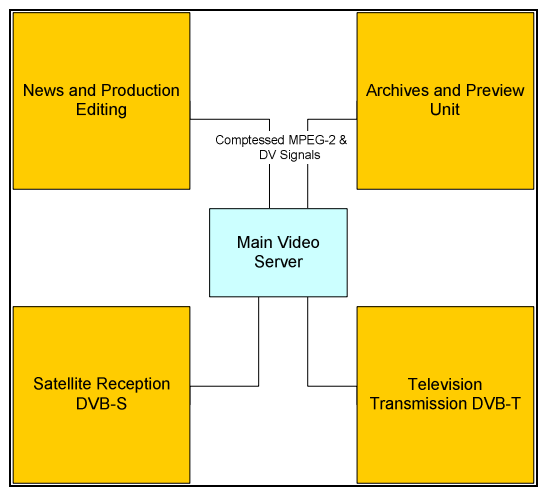

Figure 1. Block diagram of the 4 mains department networked over the main Video Server

This research is mainly concerned with audio and video signals for Television transmission, Archiving and Editing. Compressed MPEG-2 video signals are received from satellite source. The signals are demodulated using QPSK and converted to analogue before being stored in Beta Cam or DV Cam magnetic tapes. At the transmission level, analogue signals processed at the broadcast house are sent to Multicarrier Ltd (a third party organisation) for broadcast over multiples frequencies network. With the increase in number of channels, it is becoming more and more difficult for regulatory bodies to allocate bandwidth for transmission. On the other hand, digital transmission with sophisticated and robust modulation techniques is occupying limited bandwidth for the same number of transmission channels. In 2005, MBC has launched its DVB-T channels. But still, with the lack of digital networking for audio and video transmission, reference signals sent to Multicarrier Ltd, in analogue form, is processed to digital at the transmission station and aired. This results in loss in signals quality and freezing of picture frames due to high intersymbol interference (ISI) at the modulator [4].

Path loss analysis, Quality of service for the design of broadcast terrestrial network and the setting up of appropriate equipment at base station are now vital. People want to receive good quality of audio and video, mainly when the world is moving towards the digital era. Many researchers over the years since the launching of digital television have conducted research in these areas. Fernandez et al. have studied indoor digital radio reception in the Medium Wave band [8]. Heuck has derived a model of a hybrid network and performance analysis of hybrid networks is discussed [9]. In 2005, Perez-Vega and Zamanillo have done intensive research works on derivation of models and evaluation of the quality for digital broadcast network [10]. Arinda et al. [11,12] have done similar works in Spain on digital TV broadcasting. We have carried some preliminary study on field strength with few measuring data in the north [13,14, 15] and have studied the variation of the path loss using some models [16]. This paper is the continuation of research works done recently by us since 2005.

The aims of this paper are firstly to design a local area network (LAN) for multimedia capture, storage and transmission, secondly to design modulators for DVB-S at the satellite reception and DVB-T modulator for Terrestrial transmission and lastly to study the COFDM for digital DVB-T transmission for television. The objectives therefore will include optimizing the LAN network and video server, studying and comparing different modulation techniques such as QPSK, QAM64, and QAM16 and finally simulating signals for the modulators and choose for the most efficient modulator for DVB-S and DVB-T signals.

## 2. DESIGN OF A LAN FOR DIGITAL BROADCASTING

Conventional TV transmission standards are based on technology that is more than 40 years old, and the world has been split into three discrete television systems – NTSC, PAL and SECAM – that are incompatible with each other and difficult for international program exchange [2,17,18,19]. As a result there has been widespread interest in the electronics industry to develop more advanced TV systems, including digital high definition television. Digital TV refers to digital representation and processing of the signal (Digital LAN) as well as digital transmission (DVB-T) and digital reception (DVB-S) [4,7].

Digital TV broadcast involves converting images and sound into digital code. This digitization of images and sound (data) starts with compression, in order to minimize the capacity requited of transmission channels (bandwidth).Digital TV transmission standard uses processing and compression to achieve simultaneous transmission of several different television programs [5]. The quality of signal received is equivalent to the studio output.

Digital television represents a dramatic change for the production and broadcast industries, as well as the users. New technologies have recently brought tremendous flexibility in the use of different pictures format using digital compression systems. Moreover, because of digital nature of the picture information and the emergence of powerful high-speed processors and graphic cards, the computer industry is directing its main business to the broadcasting world. These converging technologies are modifying completely the existing TV environment.

## 2.1. The Need for Local Area Media Network

As discussed earlier, media network at the Mauritius Broadcasting station (MBC) concerns four main departments, namely the editing suites (News and production), preview unit, television broadcast studios and the satellite reception room. Since the creation of the station in the year 60's, heavy and bulky equipment have been utilized for image capture, processing and transmission. It's only by the end of the 90's that MBC started to invest on software based equipment and low-cost workstations for editing and transmission. Following this trend and with major changes in Broadcasting technologies (Beta cam to DV), the station has acquired numbers of workstation with powerful software (Avid non-linear editing) for editing and broadcast [18]. The editing suites are not interconnected together in a network (stand alone), and video from one editing station is copied (dubbing) on magnetic tape and transported to other departments like preview unit or television. There is loss in video quality (generation) due to numerous dubbing. The price of standard magnetic tape for video broadcast is soaring up the sky. Archiving of uncompressed videos in magnetic tape is becoming more and more difficult. Prices for Beta Cam hardware equipment are so high that maintenance of existing machines is becoming a burden. On the other hand, the prices of powerful media workstation and software are becoming more accessible. The cost of high storage hard disk (Terabytes) is much less than that of magnetic tapes of same storage capacity. For example the cost of storing 5000 hours of media on magnetic tapes is much more than that of high quality compressed media (MPEG, DV, etc) on hard disk.

The advantages of having most of the workstations grouped in a network are
- Easy sharing and transferring of data, and protection of the data
- Application sharing
- Easy interaction of other users in the network
- Sharing of peripherals devices

As shown in figures 2, 3, 4 and 5, the proposed model for the networking architecture interconnecting the four main departments inside the Broadcasting House is presented. Details of the architectures are described in the next sub-sections.

### 2.1.1. Server network connections

As shown in figure 2, the main server is a database server of around 16 terabytes of uncompressed media storage capacity. This will serve mainly the purpose of storing video files in MPEG formats as well as AVI and OMF and DV. The stored media will be used by editing suites and television for transmission. Data from satellite feed, being compressed (real time video streaming) in MPEG-2 layers are multiplexed and sent to the database on real time basis. The server is programmed in such a way that live feeds Medias from satellite can be stored for four consecutive days before being override applying a first in first out basis. The Video server (figure 2) is the heart if the network as it will communicate with all end users.MPEG-2 video files on the server will be utilized by the preview units where they are sorted and selected prior to editing and transmission.

The Video database server will be interconnected to the preview workstations, Satellite room computers, Transmission room workstations, Editing server and a Linux proxy server for internet purposes. Here also a star physical topology is adopted with a 64 ports switch connected to all the workstations and to the 8 port switch of the LANShare server throughout a gateway.

As depicted in figure 3, the gateway used can be a CISCO switch, Pix Firewall or Symantec hardware firewall with 4 ports entry [20]. This network is connected with a Symantec hardware firewall. The hardware firewall is chosen instead of the CISCO switch because of security aspect due to internet utilization. Symantec hardware firewall of same properties costs less than CISCO Pix firewall. However one advantage of Symantec firewall is that it protects individual workstation on the network from virus attack, Denial of service attack (DOS attack) privileges and can be authenticated. In practice, most computers have designed network security. Disk-based video storage is generally recognized as being more flexible than tape, particularly when considering the transfer of Data Essence and Metadata. Disks permit flexible associations of Data Essence and Metadata with the video and audio. A video server has the characteristic that it views Content from both a network / file system perspective and a video / audio stream perspective and it provides connectivity to both domains. Constraints are derived from internal server performance and from the transport / interface technology that is used when conveying the data to / from the video storage system. Due to the currently-used interfaces, there are limitations on the way that manufacturers have implemented video servers but many of these limitations will be removed as interfaces develop. The manufacturers of disk-based video storage will typically support a wide variety of formats, interfaces and Metadata capabilities. It is recommended that the manufacturer's video storage specifications include Storage capacity, Sustained data transfer, Video Essence types supported, Audio Essence types supported, Data Essence types supported, Corrected BER, Interfaces supported and Metadata and Data Essence recording limitations as the essential parameters [5].

Within the studio there are likely to be several different classes of server, corresponding to the principal work activities within the studio – in particular, play-out servers, network servers and production servers (figure 4). These will differ in storage capacity, streaming capability and video-processing capability. Here, we deal only with the issues of transportation and storage of Data Essence and Metadata. It is important that the formats of Data Essence and Metadata employed be consistent throughout the production, distribution and emission chain. It is not desirable to track identical Data Essence and Metadata with a number of different identification schemes. In the case of Metadata, a mechanism has been established through the Metadata Registration Authority. All suppliers and users are recommended to use this facility. It is further recommended that all Essence formats (Data, Video streams and Audio streams) be registered through the same authority. Users should be able to share data electronically between databases. These databases may contain either Data Essence (e.g., subtitle text) or Metadata information. Users should not be required to manually re-enter this data at any stage beyond first entry. Indeed in many instances, Metadata and some types of Data Essence ought to be automatically created. There is also a requirement for database connectivity between broadcast and business systems for the exchange of data such as scheduling information and operational results. This connectivity should be handled in a standardized way as far as possible.

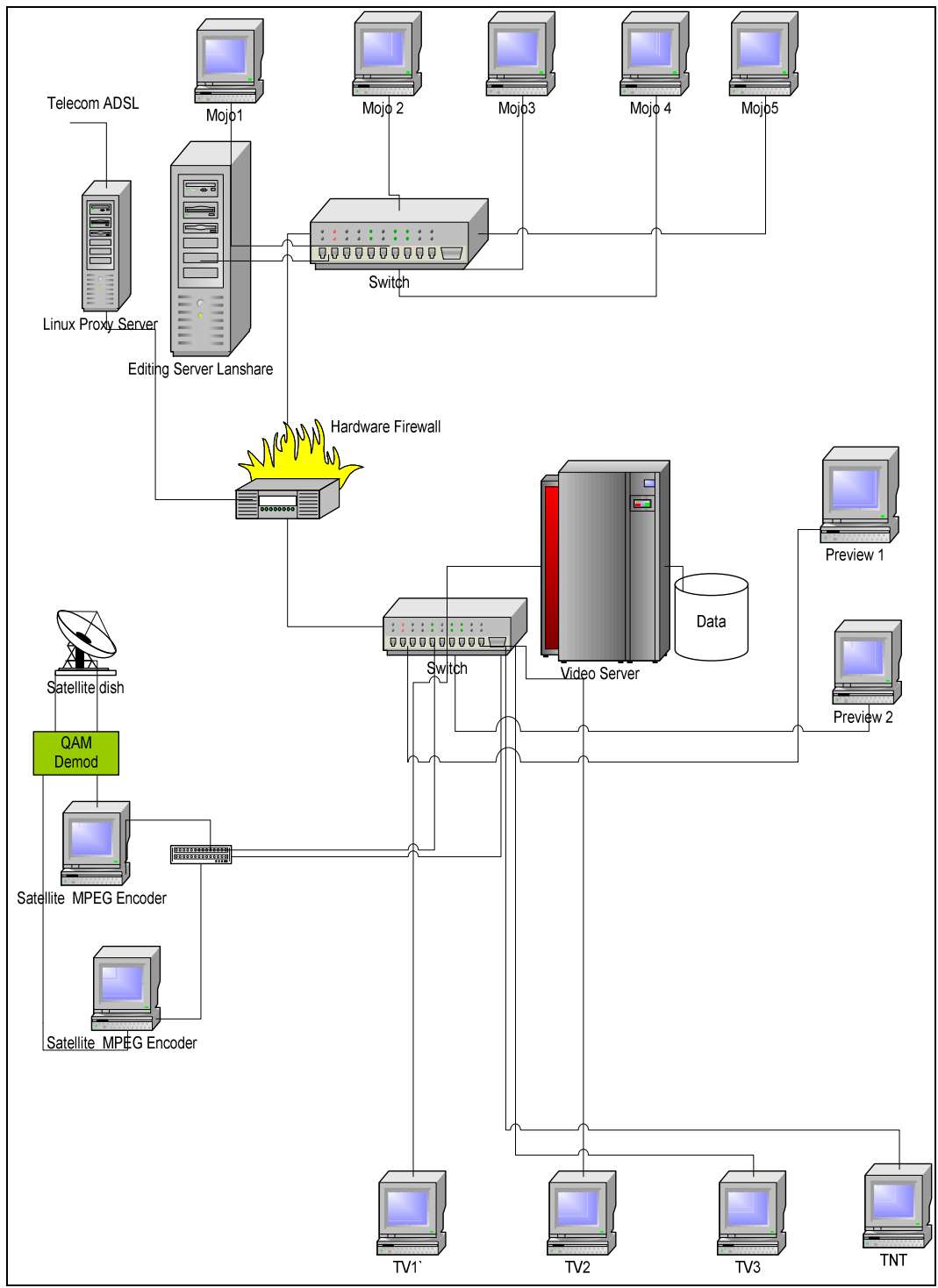

Figure 2. Proposed architecture for Digital LAN

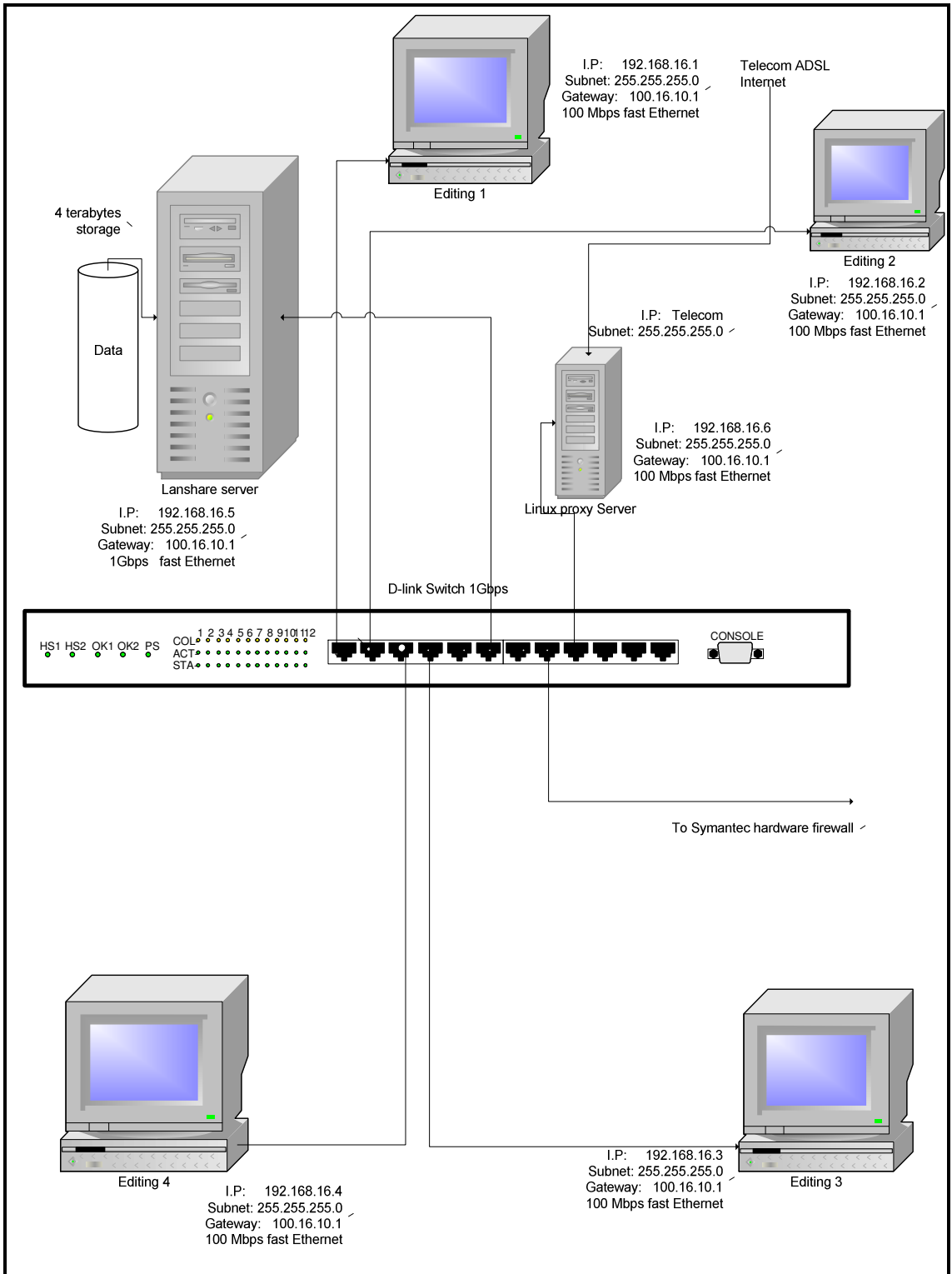

Figure 3. Detailed network design with IP addresses

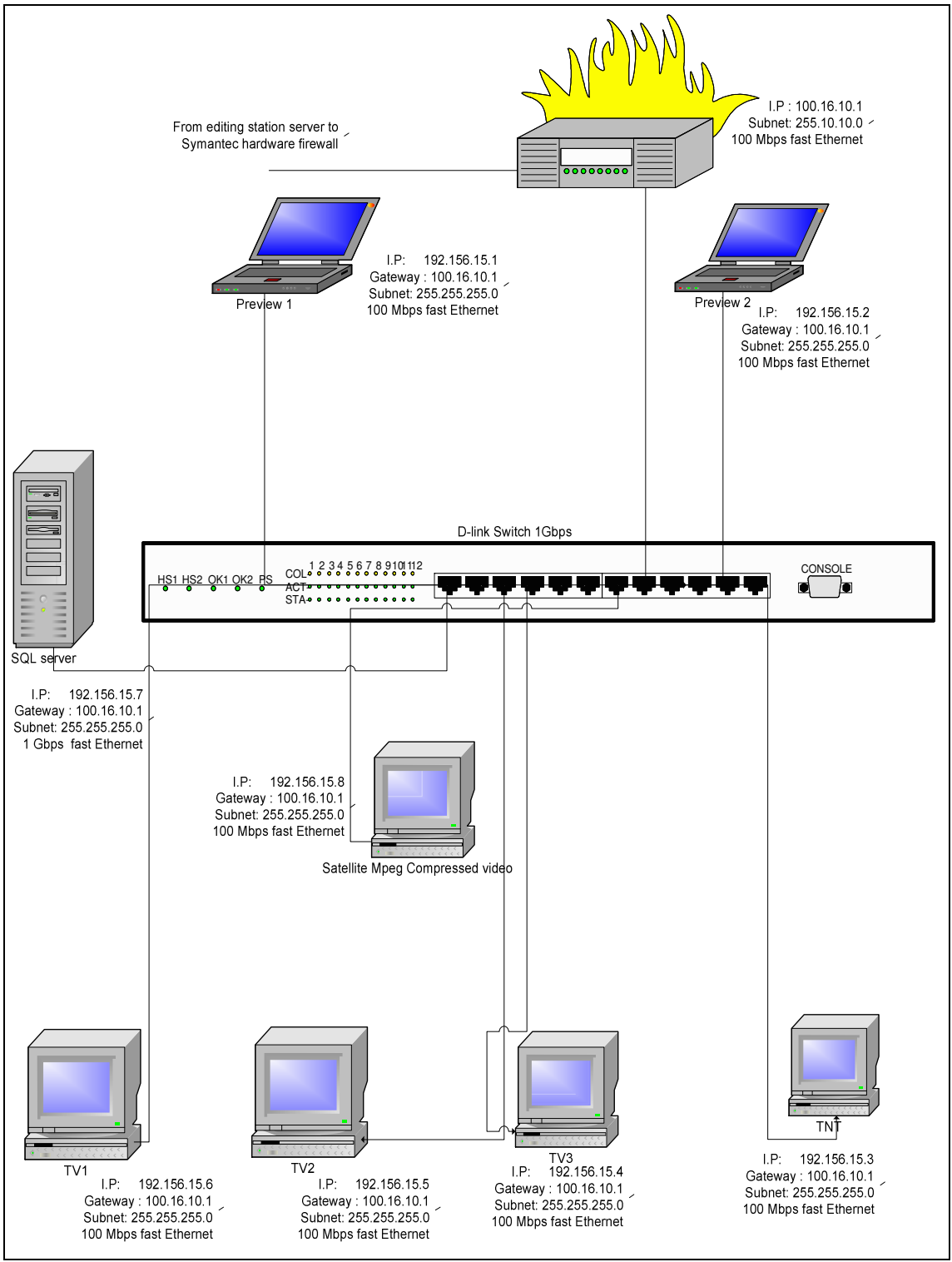

Figure 4. Detailed network design and configurations

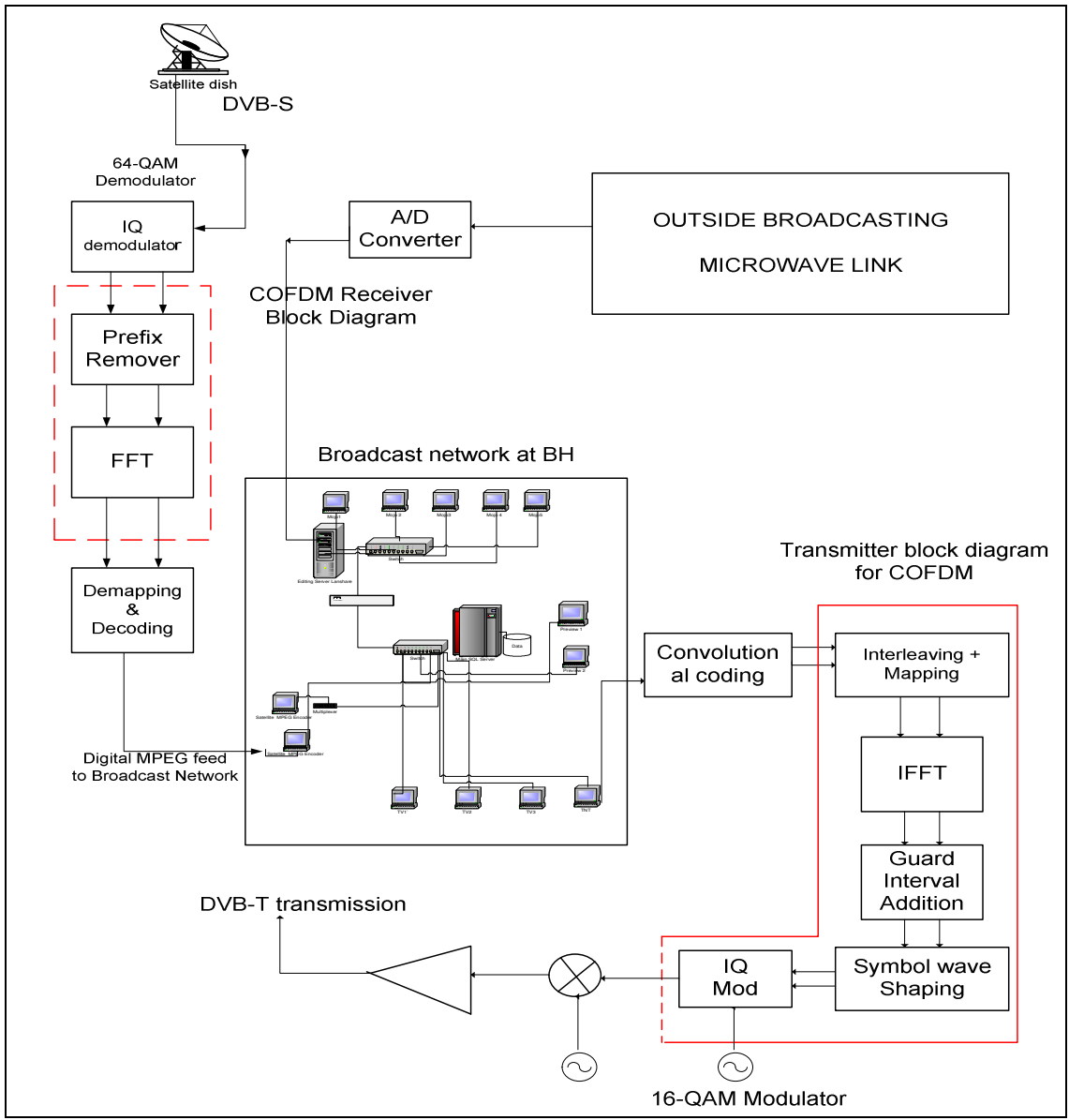

Figure 5. Entire network design and configurations

### 2.1.2 Firewall

In the network circuit designed for MBC broadcast station, a Symantec hardware firewall (figure 2) is installed and it acts as a gateway for incoming internet signals and the editing server. The firewall can be programmed to allow and block access to any workstation in the network. The firewall can be administered by an administrator using any workstation on the circuit. The hardware firewall works with a 100 Mbps network card thus allowing fluid traffic flow throughout the network. Connections to the ports are made using RJ-45 socket and 1Gbps cable having standard code IEEE 802.3. Rather than directly forwarding the incoming packets, the proxy firewall uses a special proxy application to generate fresh packets – based on the incoming request – onwards from the other network.

## 2.2. The Implementation of COFDM System for DVB-T Transmission and DVB-S Reception

As shown in figure 5, Coded Orthogonal Frequency Division Multiplexing (COFDM) has been specified for the digital broadcasting system for both audio - Digital Audio Broadcasting (DAB) - and (terrestrial) television - Digital Video Broadcasting (DVB-T) as well as satellite broadcasting (DVB-S). COFDM uses a very different method of transmission to older digital modulation schemes and has been specifically designed to combat the effects of multipath interference for mobile receivers. COFDM can cope with high levels of multipath propagation, with a wide spread of delays (Guard Spaces) between the received signals. This leads to the concept of single-frequency networks in which many transmitters send the same signal on the same frequency, generating "artificial multipath". COFDM also copes well with co-channel narrowband interference, as may be caused by the carriers of existing analogue services.

### 2.2.1. Concatenated coding system for DVB-T transmission

As depicted in figures 5 and 6, Coding of bits (Signal) is achieved into four major phases
- The input data bits are first converted to parallel by using serial to parallel block.
- Vector Reed-Solomon forward error coding is then performed
- The encoded data is then passed through a frequency interleaver where 223x8 bits symbol converted by the RS-encoder to 255-symbol are arranged in arrays where a pattern of 8-bit per block is formed.
- The signal is further encoded (inner coding) by applying fractional rate coding to further performance enhancement in the presence of frequency selective fading and interference.

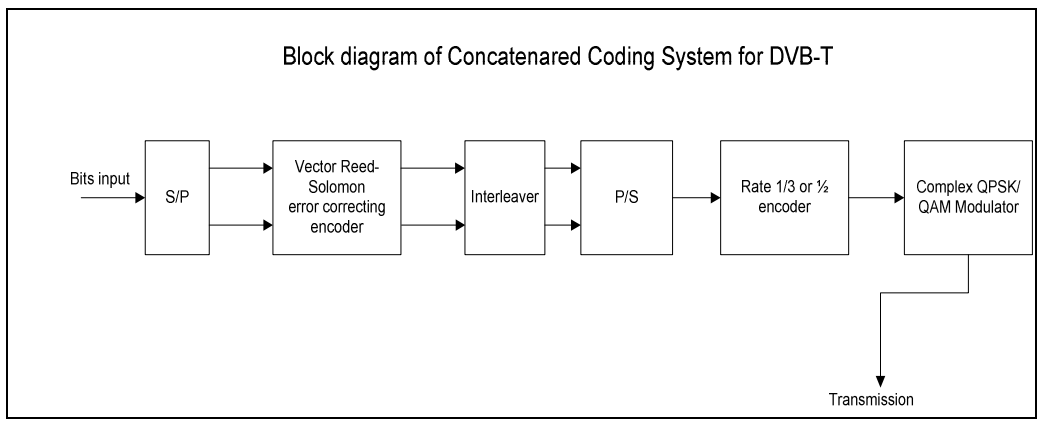

Figure 6. Concatenated coding system for DVB-T

## 2.3 Modulation Techniques

Modulation is the process of encoding information from a message source in a manner suitable for transmission. It generally involves translating a source signal (baseband) to a bandpass signal at frequencies very high compared to the baseband frequency. The bandpass signal is called the modulated signal and the baseband message signal is called the modulating signal. Modulation can be done by varying amplitude, phase or frequency of high frequency carrier in accordance with the amplitude of the message signal. The benefit of using 16-QAM or 64-QAM is that each symbol on each subcarrier can carry more bits of information. Of course, it is better to use a higher level constellation so that the overall capacity can be higher, but the drawback is that the points are

International Journal of Wireless & Mobile Networks (IJWMN) Vol. 3, No. 6, December 2011

closer together which makes the transmission less robust to errors. Fading alters both the amplitude and phase of a carrier or subcarrier, and in the mobile channel the frequency of the subcarriers are altered by a Doppler shift. However, analyzing the three modulation techniques explained, it can be said that a 16-QAM modulator having 2 subchannels and with protection distance larger than 64-QAM would perform better in the presence of frequency-selective fading environment.

### 3.4. OFDM system of transmission

The OFDM block (figure 7) is transformed into the time domain by means of IFFT (Inverse fast fourrier transform or IDFT). Guard interval is added to the signal to prevent inter symbol interference. In the receiver, the guard interval is removed and the signal is transformed into frequency domain by means of Fast Fourier transform (FFT). After demodulation bits are mapped applying soft decision values and then iterative decoding is performed by the vertebi decoder to obtain the output data.

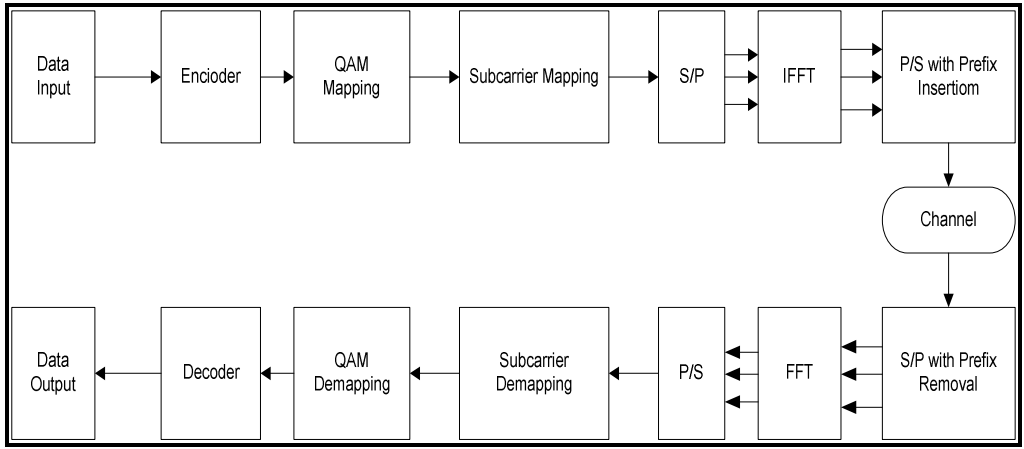

Figure 7. OFDM system of transmission

### 3.5. Transmission and reception system over designed Architecture

Figure 5 is a complete representation of the proposed digital network which encompasses the DVB-T system of transmission with a 16-QAM modulator, and a DVB-S system of satellite reception with a 64-QAM demodulator. The transmitted signal is first passed through the RS encoder for forward error correction, and then it is fractionally coded through the Vertibi encoder (addition of punctured convolution code). After interleaving the signal it is then converted into a parallel stream before being transformed into the time domain by the inverse Fast Fourier transform (IFFT). The signal is then mapped into the 16-QAM modulator with a capacity of 4 bits per symbol. After addition of guard interval; the modulated signal leaves the station as COFDM.

At the Satellite reception, the COFDM signal is digitally tracked, and then the signal is demodulated using a 64-QAM .The demodulated signal is then passed through the guard interval remover. The parallel signal is then converted back into the frequency domain by applying the Fast Fourier transform (FFT). The signal is then converted into serial form before being decoded by the Vertibi and the RS decoder.

An additional source of input to the broadcast network is the outside broadcast signal. Signal from outdoor is transmitted via microwave link; it is captured and compressed into digital form before being stored into the video server.

### 3.6. Proposed design for DVBT transmission over Mauritius

Figure 8 is therefore the proposed architecture for DVBT transmission over Mauritius. COFDM signals are modulated using QAM modulators. MPEG-2 compressed video is sent for transmission. The channel is divided into frequency spacing of 200 MHz and are orthogonally coded. The generated n random bits are encoded using a rate-3/4 punctured convolutional encoder. The resulting vector contains 4/3n bits, which are mapped to values of -1 and 1 for transmission. The puncturing process removes every third value and results in a vector of length 8/9n. The punctured code, punctcode, passes through an additive white Gaussian noise channel. After passing through a 12x12 array frequency interleaver, the signals is sent to the QAM mapper where it get modulated for transmission with appropriate guard interval.

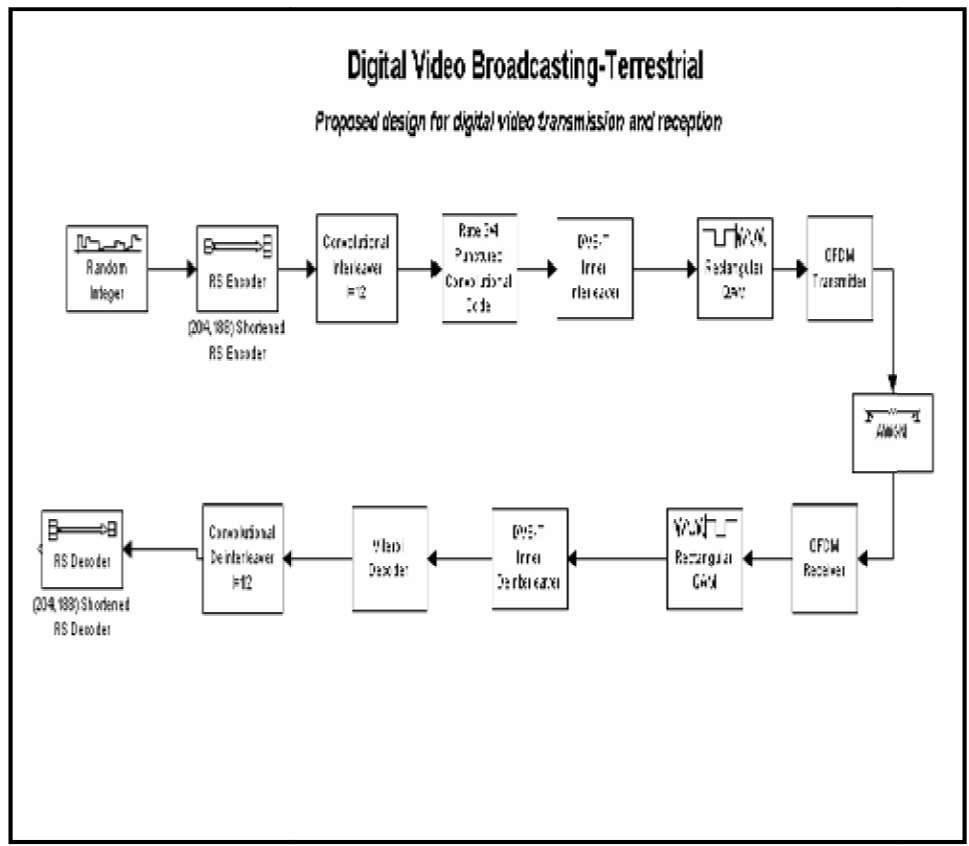

Fig.8. Proposed design for DVB-T transmission in Mauritius

## 4. RESULT AND DISCUSSION

To perform analysis QAM modulators and demodulators are simulated from MATLAB. An AWGN channel adds white Gaussian noise to the signal that passes through it. The relative power of noise in an AWGN channel is typically described by quantities such as:

(i) Signal-to-noise ratio (SNR) per sample. This is the actual input parameter to the AWGN function
(ii) Ratio of bit energy to noise power spectral density (Eb/N0).
(iii) Ratio of symbol energy to noise power spectral density (Es/N0)

## 4.1. Comparison of BER and SER for different modulators

An analysis of the bit error rate and the symbol error rate is performed while increasing the bit energy to noise power density (Eb/N0). As depicted in figures 9, 10 and 11, the graph of error probability is plotted against Eb/No (dB) for QPSK, QAM-16 and QAM-64. It is observed that as the number of bits per symbol is increased, the symbol error rate and bit error rate increase accordingly from an ordinary QPSK to QAM-64. Since the SER is more affected than the BER, it can be said that all the three modulators can be utilized for DVBT transmission at Eb/No in the range 10-15 dB. But as for QPSK, which can rotate limited amount of bits at a time, it will cause long delay in signals. QAM16 and QAM64 are therefore better for digital video broadcast.

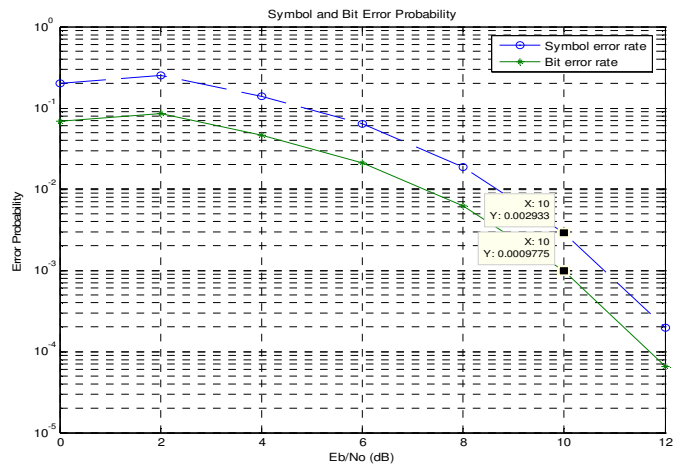

Figure 9. Graph showing symbol error rate and bit error rate for QPSK modulator

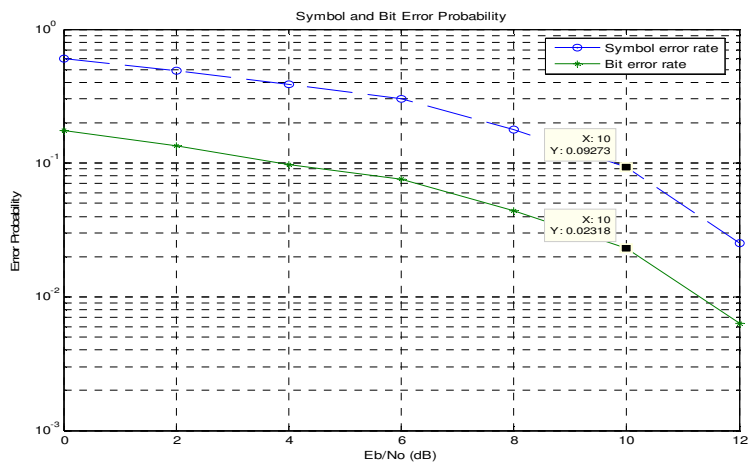

Figure 10. Graph showing symbol error rate and bit error rate for QAM-16 modulator

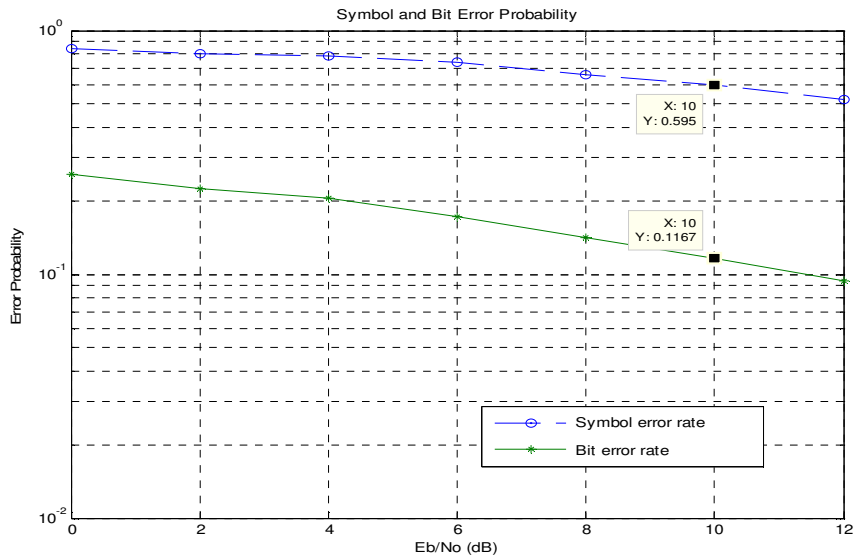

Figure 11. Graph showing symbol error rate and bit error rate for QAM-64 modulator

## 4.2. BER Tools analysis

When ratio of bit energy to noise power spectral density (Eb/N0) is set to 0-18 dB, the following outputs are observed (figure 12). Taking phase noise and AWGN into consideration and assuming an average of 15 dB of bit energy to noise density as an extreme condition, we can observe that there is a subsequent increase in the bits error rate (BER) from the QPSK model to the QAM16 and QAM64 model.

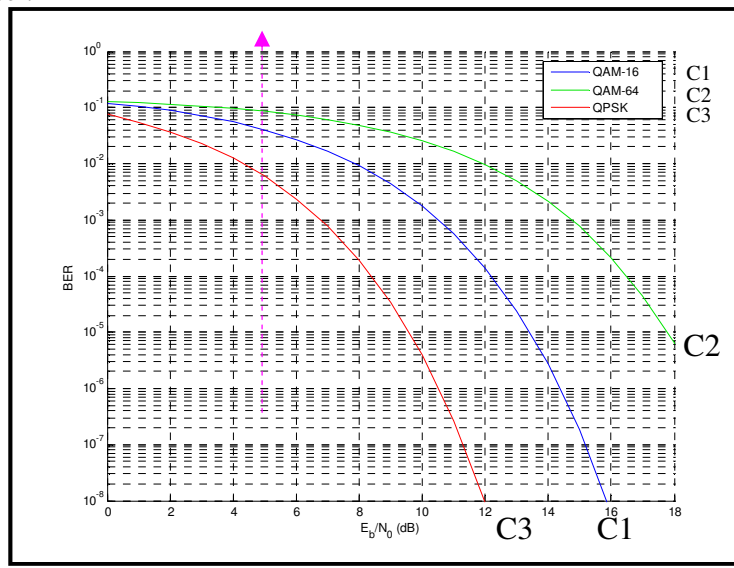

Figure 12. Graph representing the variation of BER with Eb/No

However, at 15 dB the bit error rate can still be tolerated as the respective modulators shows that the bit error rate will still be of small magnitude of $10^{-3}$. They are within the norms set up by European and American standards for broadcasting.

## 4.3. Simulations with noise addition

The main idea of this simulation is to compare two types of signal. The first one is the original signal sent for transmission and the second one is the transmitted signal. The transmitted signal is modulated, added to white Gaussian noise and phase noise, and then demodulated. The difference between the pure signal and the transmitted signal is given by the error rate calculator. Finally the total number of symbol errors is displayed.

The blocks and lines in the model describe mathematical relationships among signals and states:
(i) The Random Integer Generator block, labeled "Random Integer," generates a signal consisting of a sequence of random integers between zero and 255
(ii) The Rectangular QAM Modulator Baseband block, to the right of the Random Integer Generator block, modulates the signal using baseband QAM or QPSK modulators.
(iii) The AWGN Channel block models a noisy channel by adding white Gaussian noise to the modulated signal.
(iv) The Phase Noise block introduces noise in the angle of its complex input signal.
(v) The Rectangular QAM Demodulator Baseband block, to the right of the Phase Noise block, demodulates the signal.
(vi) In addition, the following blocks in the model help to interpret the simulation:
(vii) The Discrete-Time Scatter Plot Scope block, labeled "AWGN plus Phase Noise," displays a scatter plot of the signal with added noise.
(viii) The Error Rate Calculation block counts symbols that differ between the received signal and the transmitted signal.
(ix) The Display block, at the far right of the model window, displays the symbol error rate (SER), the total number of errors, and the total number of symbols processed during the simulation

The first simulation is performed with a 16-QAM modulator (figure 14) and demodulator units. The result obtained is tabulated (table 1) and a Q/I plot of the transmitted signal is displayed as shown in figure 13. The inter symbol interference is less in QAM-16.

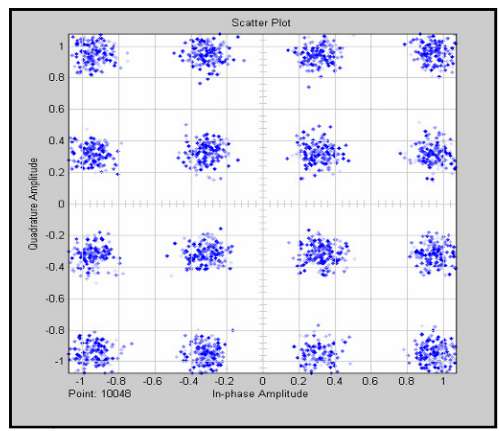

Figure 13. Scatter plot showing mapping of 10000 bits on QAM-16

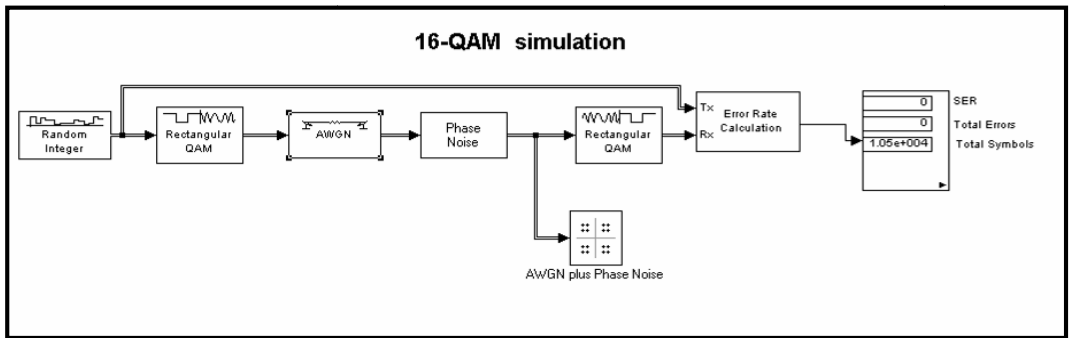

Figure 14. 16-QAM simulator

The same procedures are repeated for a 64-QAM modulators and demodulators (figure 16). The conditions of noises are kept constant. The Q/I plots of the 64-QAM and QPSK are displayed similarly (figure 15). A shown in figure 15, the inter symbol interference is more pronounced in QAM-64 compared to the result obtained as depicted in figure 13.

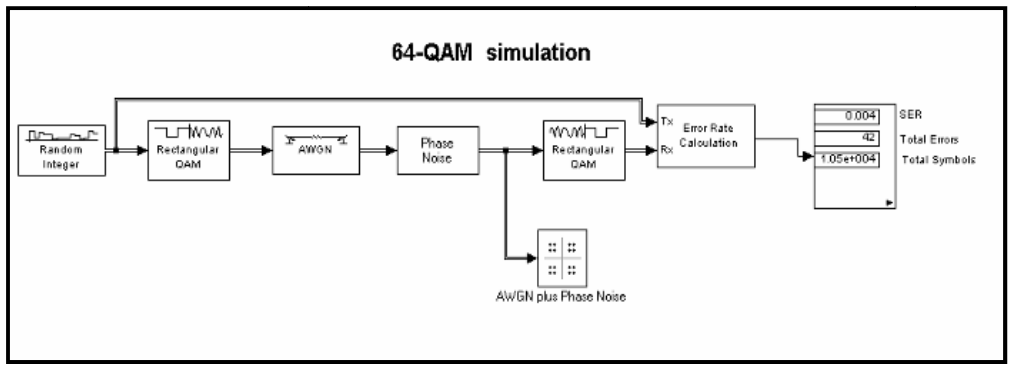

Figure 15. 64-QAM simulator

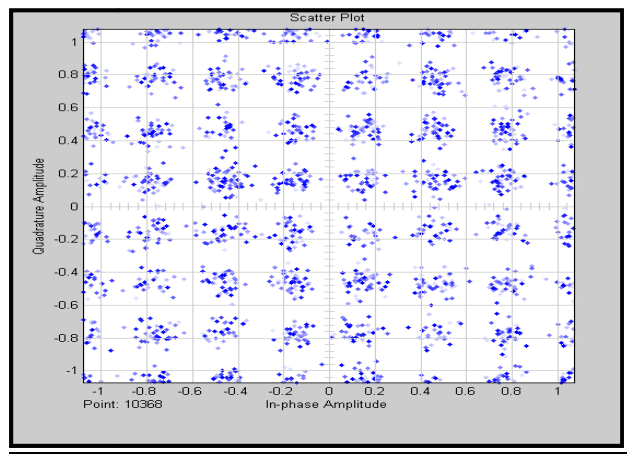

Figure 16. Scatter plot showing mapping of 10000 bits on QAM-64

Simulations are performed with the three types of modulators and the results are displayed as shown in the table below. For consistency of results, the phase noise value is kept constant at -76 dB/Hz and the total number of symbol processed is kept at 10500.The designed pi/4 QPSK simulator is also tested with 10500 symbols. The AWGN values vary from 12 dB to 15 dB and

86

results for total number of symbol error out of 10500 symbols is obtained. Symbol error rate is also tabulated for the three modulators at different AWGN values (Table 1). The values are then compared with the standard values set by the European and American acceptable norms for broadcast.

Table 1. Simulation results for QAM/QPSK modulators

| Modulator/ Demodulator | No of bits Per symbol | AWGN Eb/No(dB) | Phase Noise dB/Hz | Total Symbols | Total Errors | SER |
|---|---|---|---|---|---|---|
| QAM-16 | 4 | 12 | -76 | 10500 | 4 | 0.000381 |
| QAM-64 | 6 | 12 | -76 | 10500 | 598 | 0.05695 |
| QPSK | 2 | 12 | -76 | 10500 | 1 | 0.00095 |
| Increasing AWGN from 12 to 15 | | | | | | |
| QAM-16 | 4 | 15 | -76 | 10500 | 0 | 0 |
| QAM-64 | 6 | 15 | -76 | 10500 | 42 | 0.00400 |
| QPSK | 2 | 15 | -76 | 10500 | 0 | 0 |

Results obtained from the phase noise simulator have shown that decreasing the ratio of bit energy to noise power spectral density (Eb/N0) or decreasing the Signal to noise ratio below 15 dB increases the symbol error rate of the QAM and QPSK modulators. However, keeping the phase noise value constant at -76 dB/Hz and changing the AWGN value from 12 dB to 15 dB gives rise to a decrease in the symbol error rate for the QAM-64 modulator. Keeping the Eb/N0 at 15 dB as threshold value and increasing further by 1 dB further decreases the symbol error rate. QAM-16 performance at an ideal condition of 13 dB is optimum. The error is 1 symbol over 10500 of processed symbol. After analyzing the results, it can be said that a QAM-16 and a QPSK modulator perform better at low SNR or Eb/N0. However, the time taken to process the 10500 symbol was fastest for the QAM-64 modulator followed by the QAM-16 Modulator. QPSK with 4 symbols per rotation took much more time to process the very high number of bits.

Digital video broadcast (DVB-T and DVB-S) involve highly compressed media and high number of bits. QPSK with four symbols per mapping would definitely not process the high volume of bits from the digital channel. There will be buffering of bits at the modulator and consequently the 'gel d'image'. QAM-16 with 16 symbols per rotation and accepting 4 bits per symbol gives the ideal solution for digital video broadcast; but here also it all depends on the number of channels to be sent to the modulator for processing.QAM-64 is the easiest solution for digital video broadcast provided that the terrain condition of Mauritius is good enough. QAM-64 has been adopted by MCML (Multicarrier Mauritius Ltd) the national broadcast for the transmission of terrestrial digital video broadcasting [13]. The performance of the modulator is good in some region and at stable weather condition. Mauritius being a tropical mountainous island with fast climatic changes needs a thorough study of the whole areas. Noise variations in different regions of the island should be analysed so that a precise indication can be obtained about the efficiency of QAM-64 modulation.

Finally, as it was shown in the first simulation that symbol error rate is much higher than bit error rate, we can deduce that at 15 dB of Eb/N0 satisfactory reception of digital signal can be achieved.QAM-64 is a good solution for DVB-S and can be used to demodulate the large number of channels from the satellite reception. For a small island like Mauritius surrounded by water and mountains, QAM-16 would be the right solution for DVB-T provided that the number of transmitted channels is reasonable.

## 5. CONCLUSION

Digital transmission and single frequency network are the future of video and audio broadcasting. Most of the well known international stations in the World have migrated towards digital broadcasting. Digital broadcasting has not only cut down the cost of broadcast equipment, but it has also created lots of space over the bandwidth of transmission. This is the right time for implementing digital transmission over Mauritius. Digitalization will increase the number of broadcast channels and as well increase the quality of sound and picture that is broadcast. Digital transmission and reception requires a well designed broadcast LAN.

In this paper we have given a thorough explanation about the requirements of a video LAN. The digital network was designed according to the norms set by the European broadcasting Union. A video server was connected over a star network with the help of CISCO switch and Gateway firewall. We have given a detailed study of COFDM and modulation techniques. After the detailed design of QAM and QPSK symbol constellations, we have noted that QAM modulators and demodulators are the right solution for digital video broadcast. We found that inter symbol interference was much less in QAM-16 than the in QAM-64 and also increasing the number of symbols meant increasing the number of bits to be transmitted. Results show that the utilization of QAM-16 modulator for DVB-T transmission throughout the island would be the most viable solution. QAM-16 is robust and resists fading effects even at an AWGN value set at Eb/No of 12 decibel. Inter Symbol interference is minimal even when more that 10000 symbols are mapped. ISI for QAM-16 is within the norms set by the European Broadcasting Union. It has been demonstrated from the Phase noise simulator that out of 10 500 symbols processed at AWGN 15dB, there was no symbol error. QAM-64 demodulator would be the right solution for DVB-S since satellite reception involves large number of digital channels and require modulators with higher processing speed .However, the symbol error rate of QAM-64 is higher at AWGN 15dB and the QAM-modulator is more likely to cause ISI and ICI.  Finally, to conclude this research based on our findings it  would be recommend that we adopt COFDM as a mode of transmission and that QAM-16 modulators be used to transmit  compressed video and  in the form of MPEG-2 over Mauritius Island.

## FUTURE WORK

This research was carried out so as to develop a transmission system for digital video broadcast. Digital video broadcast is a wide field of study. Compression of signals, encoding signals and the way signals are transmitted varies a lot and can be the major field of study of any researcher. With the migration from MPEG-2 to MPEG-7 in a near future, lots of amendment will have to be made to the digital broadcast network. As an immediate future work, there is the need to study how a database system can be developed for the video server. Actually, in Mauritius, outside broadcasting is being carried out using microwave links. The future trend will be the development of an IP based transmission system over the designed digital network using fast ADSL line and router. At the transmission level further research will have to be done on the QAM modulators. We should study about possibilities of developing QAM modulators with layers of symbols. For example, QAM-16 modulator can be worked out with two layers of 16 symbols, thus allowing more channels to be processed and at the same time decreasing ICI and ISI caused by a single layer 64-QAM modulator

## ACKNOWLEDGEMENTS

The authors would like to thank all the engineers and staff of the Mauritius Broadcasting Corporation (MBC), as well as the national broadcasting television carrier, the Multi Carrier Mauritius Ltd (MCML) for their tremendous help in terms of equipment and resources.

## REFERENCES

[1]     P. Symes, (2001) "Video Compresion Demystified", Mc Graw-Hill, International Edition, ISBN 0-07-18964-5

[2]     T.S.Rappaport, (2005) "Wireless Communications Principles and practice", Prentice-Hall of India,Second Edition, ISBN 81-203-2381-5

[3]     J.H. Stott: The effects of phase noise in COFDM, EBU Technical Review, No. 276, Summer 1998.

[4]     C. Anderson & Mark Minasi, (1999) "Mastering Local Area Network", BPB Publications of India, First Indian Edition, ISBN 81-7656-060-X

[5]     ETS 300 744 (1997) "Digital broadcasting systems for television, sound and data services; framing structure, channel coding and modulation for digital terrestrial television", available at http://www.etsi.fr

[6]     Advanced digital techniques for UHF satellite sound broadcasting. Collected Papers on concepts for sound broadcasting into the 21st century, European Broadcasting Union, 1988

[7]     J.H. Stott, (1995)" The effects of frequency errors in OFDM, BBC Research and Development Report No.RD 1995/15", available at http://www.bbc.co.uk/rd/pubs/reports/1995_15.html  [] Wendell Odom, (2001) "Cisco CCNA Exam certification guide", BPB Publications of India,First Edition, ISBN 81-7635-651-4

[8]     I. Fernandez, P. Angueira, D. De la Vega, I. Pena, D. Guerra, U. Gil, (2011). "Carrier and Noise Measurements in the Medium Wave Band for Urban Indoor Reception of Digital Radio". IEEE Transactions On Broadcasting , Vol 57, Issue 4

[9]     C. Heuck, (2010) "An Analytical Approach for Performance Evaluation of Hybrid (Broadcast/Mobile) Networks", IEEE Transactions On Broadcasting, Vol. 56, No. 1, pp. 9-10

[10]    C. Perez-Vega, J.M. Zamanillo, "Path Loss Model for Broadcasting Applications and Outdoor Communication Systems in the VHF and UHF Bands", IEEE Transactions On Broadcasting, Vol. 48, No. 2, pp. 91-96, 2002.

[11]    A. Arrinda, M. Ma Velez, P. Angueira, D. de la Vega, J. L. Ordiales, Digital Terrestrial Television (COFDM 8k System) Field Trials And Coverage Measurements In Spain, IEEE Transactions On Broadcasting, Vol. 45, No. 2, pp. 171-176, (1999).

[12]     A. Arrinda, M. Ma Velez, P. Angueira, D. de la Vega, J. L. Ordiales, Local-Area Field Strength Variation Measurements Of The Digital Terrestrial Television Signal (COFDM 8k) In Suburban Environments, IEEE Transactions On Broadcasting, Vol. 45, No. 4, pp. 386-391, (1999).

[13]     V. Armoogum, T. Fogarty, K.M.S. Soyjaudah, N. Mohamudally, (2006) "Signal Strength Variation Measurements of Digital Television Broadcasting for Summer Season in the North of Mauritius at UHF Bands", Conference Proceeding of the 3$^{rd}$ International Conference on Computers and Device for Communication, pp. 89-92.

[14]     V. Armoogum, K.M.S. Soyjaudah, N. Mohamudally, T. Fogarty "Height Gain Study for Digital Television Broadcasting at UHF Bands in two Regions of Mauritius", Proceedings of the 2007 Computer Science and IT Education Conference. Nov 2007 pp. 017-025, ISBN 9789990387476**.**

[15]     V. Armoogum, K.M.S. Soyjaudah, N. Mohamudally, T. Fogarty "Path Loss Analysis between the north and the south of Mauritius with some Existing Models for Digital Television Broadcasting for Summer Season at UHF Bands", Proceedings of the 8$^{th}$ IEEE AFRICON 2007 –ISBN 0-7803-8606-X.

[16]     V. Armoogum, K.M.S. Soyjaudah, N. Mohamudally, T. Fogarty "Comparative Study of Path Loss with some Existing Models for Digital Television Broadcasting for Summer Season in the North of Mauritius at UHF Band", IEEE The Third Advanced International Conference on Telecommunications (AICT-07), ISBN 0-7695-2443-0.

[17]     EBU technical Review (2003), available at http://www.ebu.ch/en/technical/trev/trev_295-poland.pdf

[18]     Jochen Schiller, (2003) " Mobile Communications ",, Pearson Education(Singapore) Pte. Ltd, ISBN 81-7808-170-9.

[19]     Gaudrel R., Betend C., DIGITAL TV BROADCAST Field Trials on the Experimental Network of Rennes, Internal VALIDATE document from CCETT, FT.CNET/DMR/DDH, November 1997.

[20]     Wendell Odom, (2001) "Cisco CCNA Exam certification guide", BPB Publications of India,First Edition, ISBN 81-7635-651-4

**Authors**

**Sheeba Amoogum** received her BSc in Maths, Physics and Electronics in 1997 and Master in Computer Applications in 2000. She is currently doing her PhD.  She is a lecturer at the University of Mauritius.  Her fields of study are the software engineering, VoIP and networking.

**Vinaye Amoogum** received his BSc (Eng) in Computer Engineering and MSc (Eng) in System Engineering in 1995 and 1997 respectively. He completed successfully his PhD in Telecommunications in 2009.  He is currently a senior lecturer at the University of Technology, Mauritius.  His fields of study are the Telecommunications and related areas, computer science and engineering.

**Jayprakash Gopaul** is currently System Analyst/Administrator at the Mauritius Broadcasting Corporation. He has just completed an MSc degree in computational Science and Engineering in 2007.